# ExprTarget: An Integrative Approach to Predicting Human MicroRNA Targets

Eric R. Gamazon[1], Hae-Kyung Im[2], Shiwei Duan[7], Yves A. Lussier[1,3,4,5], Nancy J. Cox[1,6], M. Eileen Dolan[1,3,4]\*, Wei Zhang[8,9]\*

1 Department of Medicine, University of Chicago, Chicago, Illinois, United States of America, 2 Department of Health Studies, University of Chicago, Chicago, Illinois, United States of America, 3 Committee on Clinical Pharmacology and Pharmacogenomics, University of Chicago, Chicago, Illinois, United States of America, 4 Comprehensive Cancer Research Center, University of Chicago, Chicago, Illinois, United States of America, 5 Institute for Genomics and Systems Biology, University of Chicago, Chicago, Illinois, United States of America, 6 Department of Human Genetics, University of Chicago, Chicago, Illinois, United States of America, 7 Singapore Institute for Clinical Sciences, Agency for Science, Technology and Research (A-STAR), Singapore, Singapore, 8 Institute for Human Genetics, University of Illinois College of Medicine, Chicago, Illinois, United States of America, 9 Department of Pediatrics, University of Illinois College of Medicine, Chicago, Illinois, United States of America

## Abstract

Variation in gene expression has been observed in natural populations and associated with complex traits or phenotypes such as disease susceptibility and drug response. Gene expression itself is controlled by various genetic and non-genetic factors. The binding of a class of small RNA molecules, microRNAs (miRNAs), to mRNA transcript targets has recently been demonstrated to be an important mechanism of gene regulation. Because individual miRNAs may regulate the expression of multiple gene targets, a comprehensive and reliable catalogue of miRNA-regulated targets is critical to understanding gene regulatory networks. Though experimental approaches have been used to identify many miRNA targets, due to cost and efficiency, current miRNA target identification still relies largely on computational algorithms that aim to take advantage of different biochemical/thermodynamic properties of the sequences of miRNAs and their gene targets. A novel approach, ExprTarget, therefore, is proposed here to integrate some of the most frequently invoked methods (miRanda, PicTar, TargetScan) as well as the genome-wide HapMap miRNA and mRNA expression datasets generated in our laboratory. To our knowledge, this dataset constitutes the first miRNA expression profiling in the HapMap lymphoblastoid cell lines. We conducted diagnostic tests of the existing computational solutions using the experimentally supported targets in TarBase as gold standard. To gain insight into the biases that arise from such an analysis, we investigated the effect of the choice of gold standard on the evaluation of the various computational tools. We analyzed the performance of ExprTarget using both ROC curve analysis and cross-validation. We show that ExprTarget greatly improves miRNA target prediction relative to the individual prediction algorithms in terms of sensitivity and specificity. We also developed an online database, ExprTargetDB, of human miRNA targets predicted by our approach that integrates gene expression profiling into a broader framework involving important features of miRNA target site predictions.

**Funding:** This work was funded through the Pharmacogenetics of Anticancer Agents Research Group (http://www.pharmacogenetics.org/) by the National Institutes of Health/National Institute of General Medical Sciences (NIH/NIGMS) grant U01GM61393, National Institutes of Health/National Cancer Institute grants CA139278 and U54 CA121852, NIH/NCI Breast SPORE P50 CA125183, and the University of Chicago Comprehensive Cancer Research Center Pilot Project Program. Data deposits are supported by the NIH/NIGMS grant U01GM61374 (http://www.pharmgkb.org/). The funders had no role in study design, data collection and analysis, decision to publish, or preparation of the manuscript.

**Competing Interests:** The authors have declared that no competing interests exist.

\* E-mail: edolan@medicine.bsd.uchicago.edu (MED); weizhan1@uic.edu (WZ)

## Introduction

Gene expression is a fundamental phenotype that affects complex cellular, physiological and clinical phenotypes including disease risk as well as individual response to therapeutic treatment. For example, gene expression alterations have been implicated in the etiologies of common diseases such as cancers [1–3], cardiovascular diseases [4], and psychiatric disorders [5]. Previous studies using the International HapMap Project [6,7] lymphoblastoid cell lines (LCLs) derived from individuals of European (CEU: Caucasians from Utah, USA), African (YRI: Yoruba people from Ibadan, Nigeria) and Asian (CHB: Han Chinese from Beijing, China; JPT: Japanese from Tokyo, Japan) ancestry have shown that common genetic variants including single nucleotide polymorphisms (SNPs) and copy number variants (CNVs) account for a substantial fraction of variation in gene expression within a population and between populations [8–14]. Furthermore, pharmacogenomic studies based on these HapMap cell lines [6,7] strongly suggest that response to therapeutic treatment is likely to be a complex phenotype affected by genetic factors that alter gene regulation, especially in the form of eQTLs (expression quantitative trait loci) [15–17].

In addition to eQTLs, more recently, microRNAs (miRNAs) (~700 known in humans to date), a family of small (21–23 nucleotides), single-stranded, non-coding RNAs, have been shown to be an important class of gene regulators that generally down-regulate gene expression through sequence-specific binding to the 3′ untranslated regions (UTRs) of target mRNAs [18]. In humans, miRNAs are predicted to potentially target up to one third of protein-coding genes [19]. Global microRNAome profiling has demonstrated significant changes in the expression of multiple miRNAs in a growing list of human diseases, including neurodegenerative diseases [20], heart diseases [21] and cancer [22]. Because an individual miRNA may regulate multiple mRNAs, a comprehensive and reliable catalogue of miRNA gene targets should enhance our understanding of the complexity of gene regulatory networks in a cell, as well as facilitate the shifting of focus from miRNA gene identification to functional characterization. Due to the lack of high throughput experimental technique for identifying the targets of miRNAs, only a small proportion of the targets of the potentially more than 1000 human miRNAs – such as those in TarBase [23,24] (a manually curated database of experimentally supported miRNA targets) – have been confirmed experimentally. Therefore, several computational and bioinformatic approaches have been developed for large-scale prediction of miRNA targets including miRanda [25,26] (based on sequence complementarity, free energy of the RNA-RNA duplex, extent of conservation), TargetScan [27–29] (based on seed complementarity, thermodynamic free energy of binding, conservation over different species), and PicTar [30] (based on seed complementarity, thermodynamics and a combinatorial prediction for common targets in sets of coexpressed miRNAs). The various computational prediction algorithms have been found to suffer from significant false positive and false negative rates. For example, the miRanda algorithm [25,26] was estimated to have a high false-positive rate at 24–39% in an early study [31]. A high false-negative rate is also expected for the current miRNA target prediction programs, largely due to their requirements for evolutionary conservation, while many of the experimentally-supported targets (e.g., those from TarBase [23,24]) may not be conserved in other species [32]. Though using combinations of two or more of these computational approaches may improve the sensitivity or specificity of the prediction, the degree of overlap of their predictions is poor [32]. To date, a few algorithms have been developed to leverage existing approaches involving relevant miRNA binding site considerations such as thermodynamics, sequence complementarity, conservation, and gene expression profiles [32]; however none of these algorithms provide a genome-wide map of predictions that utilizes post-transcriptional regulation within a broad framework of relevant target site prediction features as well as utilizes the experimentally validated binding sites as training set.

Though experimental testing is critically important for validating any putative miRNA targets, we propose here a novel bioinformatic approach, ExprTarget (**Fig. 1**), to predicting human miRNA targets by integrating select computational algorithms and our recently-generated miRNA expression data and previously published mRNA data [33] on 58 unrelated HapMap CEU LCLs [6,7]. We demonstrate that a significant improvement of performance in terms of sensitivity and specificity can be achieved by integrating both the computational algorithms as well as the experimental miRNA expression data. Furthermore, the study we report here of miRNA-mRNA relationships in the HapMap samples [6,7] extends, in the direction of miRNA-mediated gene regulation, earlier studies on these same samples that have successfully been used as models for studies of complex traits [34] and for pharmacogenomic studies [15–17]. We developed an online resource, ExprTargetDB (http://www.scandb.org/apps/microrna/) to: 1) enable user-friendly queries of a comprehensive catalogue of miRNA targets using this integrative approach; 2) provide a reference dataset of miRNA-mRNA relationships on the HapMap samples and; 3) advance our understanding of gene

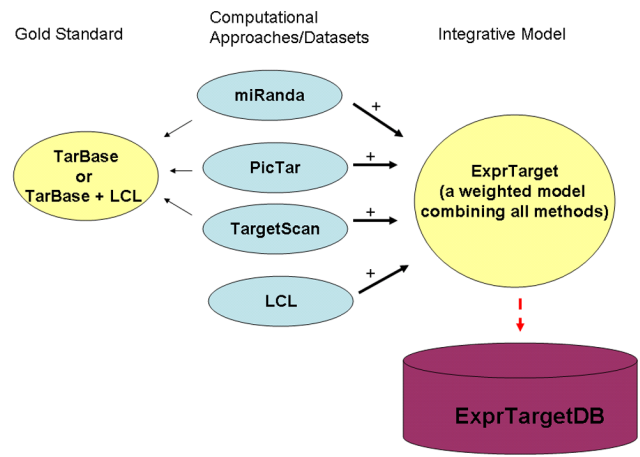

**Figure 1. ExprTarget integrates various methods and datasets.** Individual computational methods (miRanda, PicTar, TargetScan) were evaluated using both TarBase and (TarBase + LCL) as gold standards. ExprTarget integrates individual computational methods (miRanda, PicTar, TargetScan) and the LCL expression data. ExprTarget was evaluated using TarBase as gold standard. ExprTargetDB was developed to house the predictions by ExprTarget. LCL refers to the miRNA and mRNA expression data generated on a panel of lymphoblastoid cell lines from the HapMap.
doi:10.1371/journal.pone.0013534.g001

regulatory networks and of the contribution of miRNAs to complex traits.

## Results

### Pair-wise comparisons between prediction algorithms

Pair-wise comparisons show that the miRNA targets predicted by different algorithms are generally not correlated (**Fig. 2A, 2B, 2C**) in the sense that good candidates for one algorithm do not tend to be good candidates for the other algorithms. PicTar [30] scores, TargetScan [27–29] scores and miRanda [25,26] scores are not correlated for the same miRNA targets predicted by these algorithms. We compared the distribution of experimentally-validated targets (from TarBase [23,24]) using the targets' miRanda [25,26] scores (**Fig. 2D**), PicTar [30] scores (**Fig. 2E**) and TargetScan [27–29] scores (**Fig. 2F**) with the corresponding overall distribution of scores. This analysis enables us to compare, for each score bin defined by an algorithm, the proportion of target site predictions and the proportion of experimentally validated predictions falling within the bin; particularly, it shows the proportion of experimentally supported targets that overlap with the highest scoring bins.

### Performance of individual prediction algorithms

Individual prediction algorithms were evaluated for performance (see Materials and Methods) using either TarBase [23,24] alone or TarBase [23,24] combined with the expression-corroborated miRNA targets (TarBase+LCL) as gold standard. The receiver operating characteristic (ROC) curves, which plot the true positive rate (sensitivity) and false positive rate (1-specificity), for the individual algorithms are shown in **Fig. 3**. Each point on the ROC curve is a specificity/sensitivity pair corresponding to a score threshold. A comparison of the diagnostic performance of the three prediction algorithms using only TarBase [23,24] as gold standard is shown in **Fig. 3** (**Fig. 3A, 3B, 3C**). The same analysis was repeated using (TarBase+LCL) as gold standard to gauge the effect of the choice of gold standard; the results for the three prediction approaches are shown in **Figure S1**.

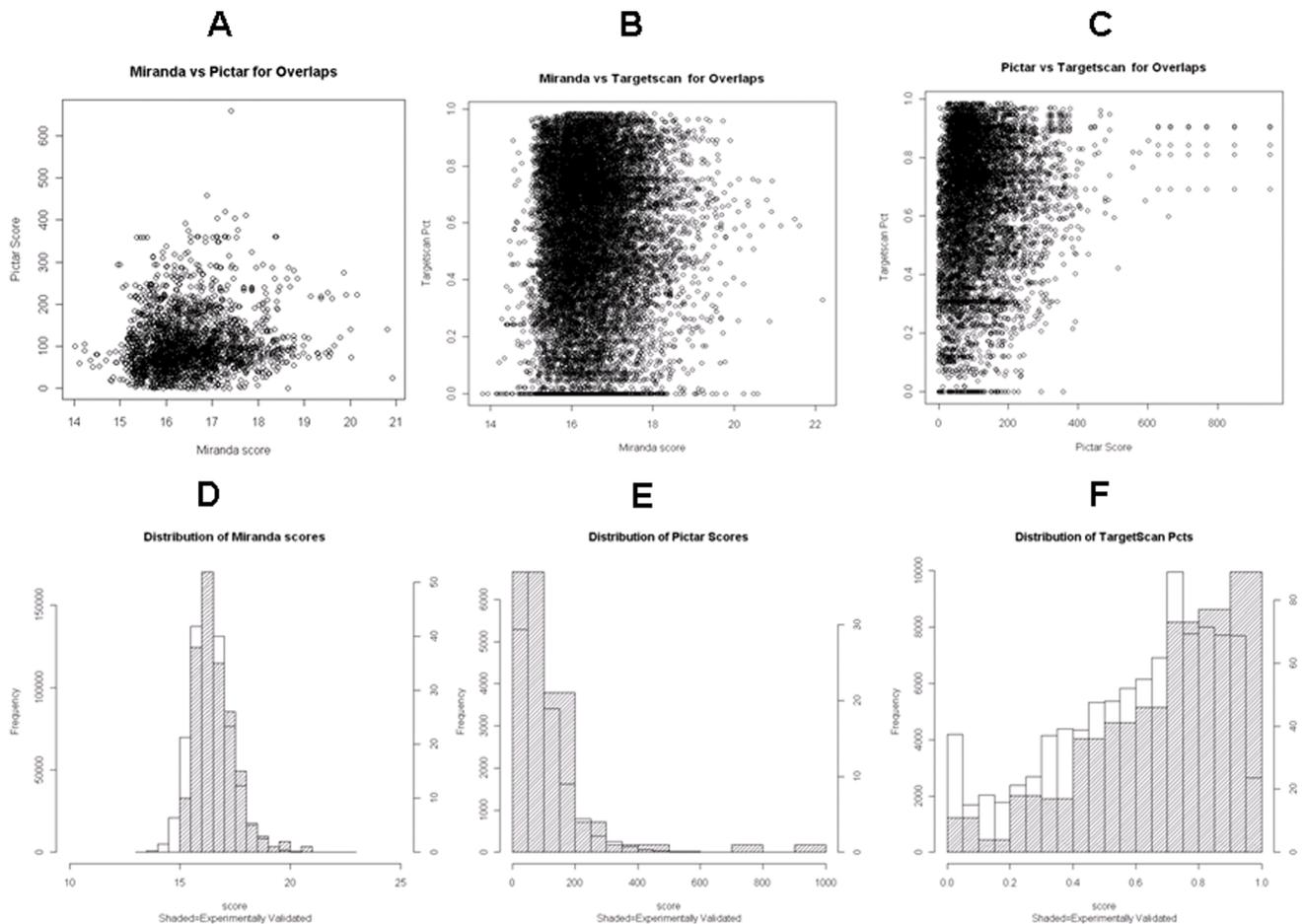

**Figure 2. Prediction results from existing computational approaches are not correlated.** Prediction scores (TargetScan, PicTar, and miRanda) for the same miRNA targets are plotted to show pair-wise comparisons (A, B, C). The distributions of scores for targets from the TarBase (experimentally-validated) are shown with the scores for the full set of targets from individual prediction algorithms (D, E, F). (A) miRanda (x-axis) scores are plotted against PicTar scores (y-axis); (B) miRanda (x-axis) scores are plotted against TargetScan scores (y-axis); (C) PicTar scores (x-axis) are plotted against TargetScan scores (y-axis); (D) Histogram of experimentally-validated targets with the distribution of miRanda scores (left y-axis is for the miRanda p values; right y-axis is for the TarBase targets); (E) Histogram of experimentally-validated targets with the distribution of PicTar scores (left y-axis is for the PicTar scores; right y-axis is for the TarBase targets); and (F) Histogram of experimentally-validated targets with the distribution of TargetScan scores (left y-axis is for the TargetScan scores; right y-axis is for the TarBase targets).
doi:10.1371/journal.pone.0013534.g002

### Integrating different algorithms and miRNA expression in miRNA target prediction

By incorporating the computational approaches included in this study (miRanda [25,26], PicTar [30] and TargetScan [27–29]) as well as the expression-based prediction model into an integrative model ExprTarget based on sigmoidal modeling (see Materials and Methods), we observed that the performance in terms of sensitivity and specificity is greatly improved using TarBase [23,24] as gold standard (**Fig. 4**). A similar ROC-based performance evaluation shows that ExprTarget is a much better classifier in discriminating the experimentally verified targets from the non-experimentally supported ones than the random guessing procedure indicated by the line of no-discrimination (**Fig. 4**) or, indeed, any of the existing computational solutions (**Fig. 3**) evaluated in this study. Reduced models (e.g., the combination of two individual methods as well as the expression-based approach) lead to decreased predictive performance, as measured by the area under the ROC curve.

Restricting the gold standard to the subset of TarBase [23,24] that excludes the high-throughput assays, we observed that the improvement in predictive performance for ExprTarget relative to the individual methods continues to hold robustly (**Figure S2**). From each computational algorithm incorporated by ExprTarget, a parameter that quantifies the confidence of each binding site is used so that the final model is target-site based. The use of a score from the expression data that is gene-based rather than target-site based (e.g., when the score from the expression data is defined as the minimum of all p values for the gene) demonstrates the robustness of our primary finding, namely, incorporating these algorithms through sigmoidal modeling improves predictive performance (**Figure S3**). To evaluate whether the fitted model can be generalized to as-yet-unseen data (given that ExprTarget uses the experimentally validated targets as training set), we proceeded to do cross-validation (see Materials and Methods) on ExprTarget. Cross-validation enabled us to evaluate how well our predictive model, which was defined by the use of training data, would perform on future data. Using 10-fold cross validation, ExprTarget resulted in a mean prediction error of 0.000277.

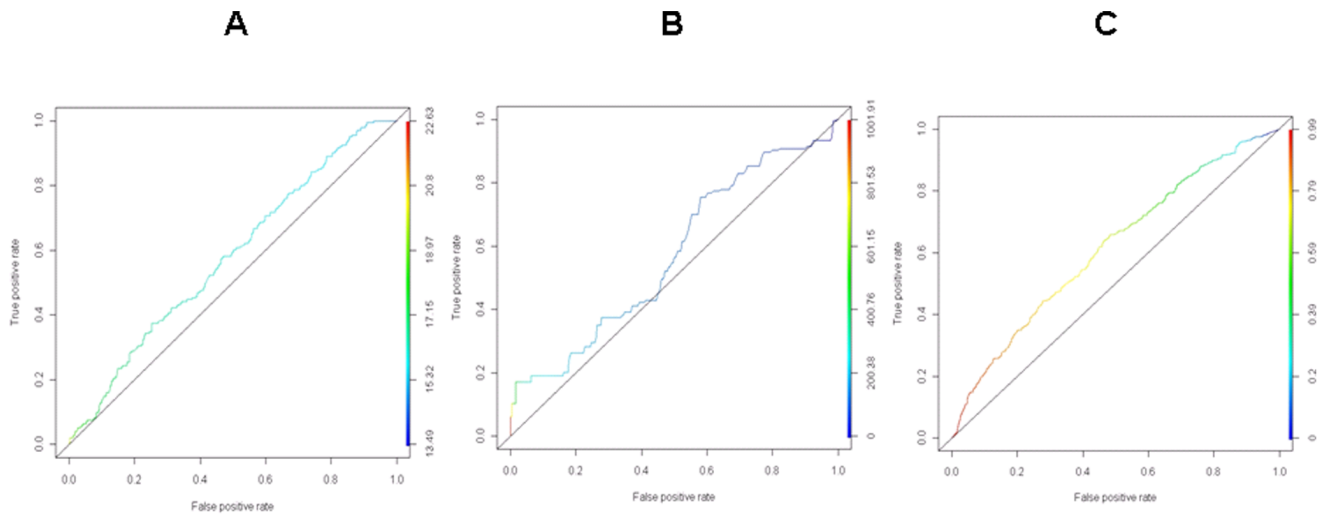

**Figure 3. Individual performance of foundational prediction algorithms.** The three prediction algorithms were evaluated using ROC curves, which plot the true positive rate (sensitivity) and the false positive rate (1-specificity) at various score thresholds. TarBase was used as gold standard. The line of no-discrimination was drawn from the left bottom to the top right corners. (A) miRanda vs. TarBase; (B) PicTar vs. TarBase; and (C) TargetScan vs. TarBase.
doi:10.1371/journal.pone.0013534.g003

### ExprTargetDB, a database for human miRNA targets

We developed a web-based database, ExprTargetDB (http://www.scandb.org/apps/microrna/), to provide user-friendly queries of our miRNA-mRNA association data in the context of other computationally-predicted miRNA targets including miRBase [25,26], TargetScan [27–29] and PicTar [30] as well as the experimentally-supported miRNA targets [23,24]. The miRNA expression profiling is, to our knowledge, the first such reported dataset on the HapMap samples. Though the initial dataset was generated from the samples of European descent only, ExprTargetDB will hold the results assayed from other HapMap populations. Since previous studies on the HapMap samples have shown that complex traits [34] and pharmacologic phenotypes [15–17] are affected by genetic variants that alter gene regulation, this dataset on miRNA-mediated gene regulation should be a tremendous resource to the scientific community. ExprTargetDB supports downloading of the complete list of expression-corroborated miRNA targets. In addition, ExprTargetDB can be queried by either miRNA or gene target, in single or batch mode. A query of miRNA should utilize the nomenclature of the miRBase [25,26] (e.g., "hsa-miR-138"), while a query for gene target should use the official gene symbols (e.g., "PAPD5"). A successful search outputs a table of miRNA targets (miRNA-centric) or miRNAs (gene target-centric). A link to our SCAN database (SNP and CNV Annotation, http://www.scandb.org/) [35] is also provided for more information on eQTLs of the gene targets to supplement the information on miRNA-mediated gene regulation. A search example is discussed in the online tutorial.

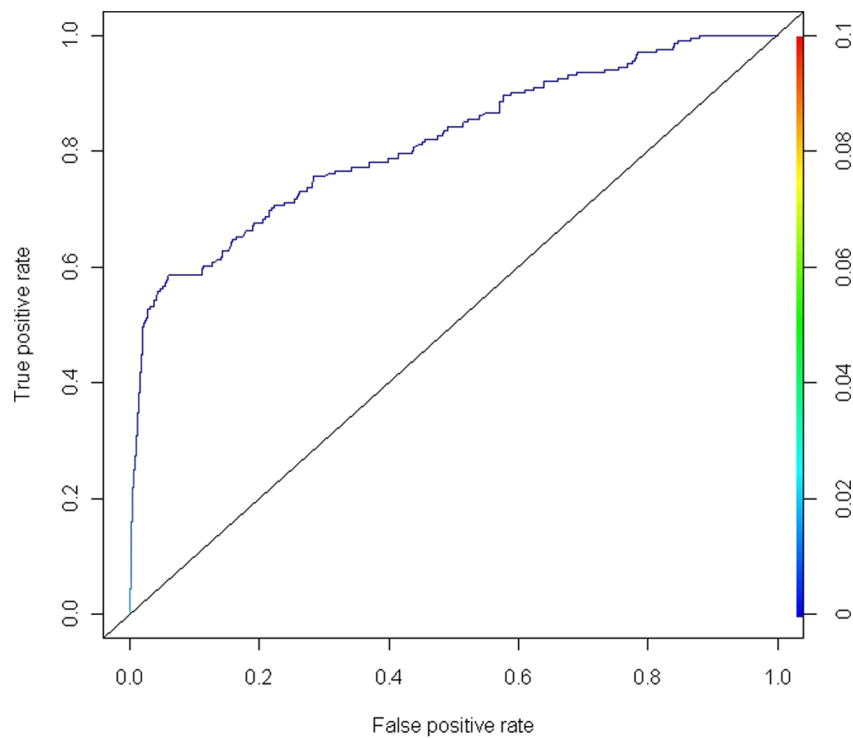

**Figure 4. Performance of ExprTarget, an integrative prediction algorithm.** ExprTarget was assessed by plotting the ROC curve, which shows the true positive rate (sensitivity) and the false positive rate (1-specificity) at various thresholds. The database of manually-curated experimentally verified targets, TarBase, was used as gold standard. The line of no-discrimination was drawn from the left bottom to the top right corners.
doi:10.1371/journal.pone.0013534.g004

## Discussion

The identification of the transcript targets regulated by miRNAs promises to greatly enhance our understanding of gene regulation and to provide important insights into the genetic/epigenetic basis of various human diseases such as those that have been found to be associated with altered or abnormal miRNA expression [20–22]. As a result of the multiplicity of computational algorithms that have been developed, two distinct problems arise: (1) the problem of inter-method reliability, and (2) the problem of integrating the results obtained from the various methods into a single optimal score. The present study conducted a comparative analysis of the most frequently used target prediction algorithms and developed an integrative approach, with certain analytically attractive properties, that improves the predictive performance in relation to the foundational prediction methods.

The problem of inter-method reliability is concerned with the degree of agreement between the various computational methods and with the predictive performance of each algorithm. Many available computational prediction methods take advantage of the biochemical or thermodynamic properties of the binding between miRNAs and their cognate mRNA transcripts. For the three frequently used algorithms we tested (i.e., miRanda [25,26], PicTar [30], TargetScan [27–29]), pairwise comparisons reveal that their prediction results are generally not correlated. This unanticipated lack of correlation between these prediction algorithms presumably reflects the differing emphasis on biochemical/thermodynamic factors as well as the extent of the use of evolutionary conservation. Surprisingly, we found very different distributional patterns for the existing prediction approaches when comparing the experimentally-validated miRNA targets (from TarBase [23,24]) predicted by an algorithm with the full set of targets for the same algorithm. For example, **Fig. 2D** shows that the distribution of miRanda [25,26] scores for predicted targets is similar to the distribution of miRanda [25,26] scores for only the experimentally supported targets; each is bell-shaped and peaks in the same score bin. In contrast, the distributions of the scores from the other two methods show quite distinct patterns.

An ROC analysis using TarBase [23,24] on each of the existing computational methods would seem to suggest that TargetScan [27–29] may yield slightly better performance than the other computational methods, based on the area under the ROC curve (AUC). However, caution must be exercised in the interpretation of these results. We sought to evaluate the effect of the choice of gold standard on the performance evaluation. A comparative ROC analysis for the three methods, using either TarBase [23,24] or (TarBase+LCL), shows the dependence of the performance evaluation on the use of a different gold standard. TarBase [23,24] is a manually-curated database with results that assume particular prediction approaches and that were derived during the experimental verification of a prediction algorithm; on the other hand, it has also been shown that a substantial proportion of TarBase predictions are non-conserved target sites and would not have been predicted by computational approaches that assume evolutionary conservation [32]. Due to these biases, it is informative to assess the performance of the individual computational approaches using another benchmark. The miRNA-mRNA associations generated from our HapMap LCL [6,7] data are genome-wide, facilitate a comparative analysis, and may serve as a reference dataset on the regulatory effects of miRNAs. However, this gold standard could reflect the possible biases from the high-throughput expression microarrays.

The problem of integrating existing computational approaches is concerned with identifying a parsimonious model, which utilizes the accumulated knowledge base of experimentally supported miRNA target sites in defining the training algorithm. The parsimonious model is selected from a potential hierarchy of models derived from various combinations of existing algorithms. ExprTarget is inspired by a multivariate logistic regression model for a binary outcome and predictor variables from existing computational solutions. ExprTarget has certain attractive features besides the relevant criterion of parsimony. In relation to existing computational methods, ExprTarget shows a greater predictive power to discriminate the experimentally verified targets based on ROC curve analysis. To gauge the accuracy, we conducted $K$-fold cross-validation ($K = 10$) on our proposed algorithm, resulting in a mean prediction error of 0.000277. There are, of course, other ways of combining individual computational approaches such as taking various unions or various intersections of the corresponding result sets. However, while unions of computational approaches may achieve a higher level of sensitivity than the individual approaches, this gain comes at the cost of a reduction in specificity. In the same vein, while intersections of computational approaches may achieve a higher level of specificity, they also generally achieve a much reduced sensitivity [32]. Alternatively, non-linear statistical models may provide an excellent fit; however, the parameters in these models are often difficult to interpret. The application of a complex model with many degrees of freedom to produce a satisfactory fit may come at the cost of failure to replicate in as-yet-unseen data. The model we propose in ExprTarget enables us to evaluate the relative importance of the predictors in miRNA target site prediction.

We acknowledge that miRNA-mediated regulation of gene expression may be tissue-specific and population-specific. Our miRNA expression data on LCLs represent one tissue type in one population (the CEU samples); therefore, it is likely that we might miss some information on miRNAs not expressed in these samples or differentially expressed between human populations. Nevertheless, the approach we propose here is completely extensible. With the availability of more data sets on other tissues (e.g., liver) and the development of new methods that may utilize novel characteristics of the miRNA-mRNA relationships, ExprTarget is amenable to refinement to guide the prediction of miRNA targets.

Finally, the prediction results of our integrative approach are provided through an online searchable database, ExprTargetDB, which can also be used to compare the prediction results (i.e., p values, scores) from the various computational solutions we tested in this study (i.e, miRanda [25,26], PicTar [30], TargetScan [27–29], TarBase [23,24], and our LCL data). In particular, the expression dataset in ExprTargetDB constitutes the first reported study of miRNA-mediated gene regulation using the HapMap samples [6,7] that have been so important in transcriptome-based studies of complex traits [34] and pharmacologic phenotypes [17]. Given the role of miRNAs on the regulation of expression, we also provide additional relevant information (e.g., eQTL annotation through the SCAN database [35]) in ExprTargetDB to inform studies of the complex networks of gene expression. ExprTargetDB has been designed to seamlessly integrate into existing genomic and pharmacogenomic resources (e.g., PharmGKB [36]).

## Materials and Methods

### MiRNA profiling of unrelated HapMap LCLs

MiRNA expression was measured in unrelated HapMap LCLs [6,7] including 58 CEU (Caucasians from Utah, USA) samples using the Exiqon miRCURY$^{TM}$ LNA Array v10.0 (~700 human miRNAs, updated to miRBase 11.0 annotation [25,26]) (Exiqon, Inc., Denmark). The HapMap cell lines [6,7] were purchased from Coriell Institute for Medical Research (Camden, NJ). The details for cell line culture were described in a previous publication [10]. Total RNA was extracted using miRNeasy Qiagen Kit (Qiagen, Inc., Valencia, CA) according to manufacturer's protocol. Array hybridization was performed by Exiqon. The quantified signals were background corrected using normexp with offset value 10 based on a convolution model [37] and normalized using the global Lowess regression algorithm. In total, 225 miRNAs were found to be expressed in these samples (unpublished data).

### Expression-supported human miRNA targets

Associations between the 225 expressed miRNAs and potential gene targets were evaluated using the miRNA profiling data and our previously-generated Affymetrix Human Exon 1.0ST array data on the same cell lines (~10,000 mRNA transcripts with reliable expression) (NCBI Gene Expression Omnibus Accession: GSE9703) [33]. Association analyses in the CEU samples were carried out using the lm function of the R Statistical Package [38].

### Computationally-predicted miRNA targets

1. The miRBase Targets Release Version v5 (http://www.ebi.ac.uk/enright-srv/microcosm/htdocs/targets/v5/) (i.e., microCosm) provides computationally predicted targets for miRNAs using the miRanda algorithm [25,26], which uses dynamic programming to search for maximal local complementarity alignments corresponding to a double-stranded anti-parallel duplex. miRanda [25,26] also takes into account the extent of conservation of the miRNA targets across related genomes. The miRBase Targets Release Version v5 is comprised of gene target predictions for 711 human miRNAs.

2. TargetScan (http://www.targetscan.org/) (TargetScanHuman Release 5.1, April, 2009) [27–29] predicts miRNA gene targets by searching for the presence of conserved sites that match the seed region of each miRNA. Predictions are ranked using site number, site type, and site context, which include factors that influence target-site accessibility.

3. PicTar [30] (http://pictar.mdc-berlin.de/) takes sets of coexpressed miRNAs and searches for combinations of miRNA binding sites in each 3′UTR. Like TargetScanS [27–29], PicTar also requires target conservation across several species [30]. PicTar has target prediction information for human miRNAs based on conservation in mammals (human, chimpanzee, mouse, rat, dog) [30].

### Experimentally-supported human miRNA targets

TarBase [23,24] (http://diana.cslab.ece.ntua.gr/tarbase/), which houses a manually curated collection of experimentally tested miRNA targets in a variety of species including human, mouse and several other model organisms. Each target site is described by the miRNA that binds it, the gene in which it occurs, the nature of the experiments that were conducted to test it, the sufficiency of the site to induce translational repression and/or cleavage, and the paper from which all these data were extracted. The current TarBase [23,24] v.5c (June, 2008) covers 1122 distinct miRNA-gene pairs for 143 human miRNAs.

### Comparing computationally-predicted and experimentally-supported miRNA targets

1. Pair-wise comparisons of miRanda scores [25,26], PicTar scores [30] and TargetScan scores [27–29] were performed to evaluate the correlations between the computational algorithms. Predicted miRNA targets overlapping between every pair were included in this analysis: 21590 between miRanda [25,26] and TargetScan [27–29], 2465 between miRanda [25,26] and PicTar [30], and 8707 between TargetScan [27–29] and PicTar [30].

2. The distribution of experimentally-supported miRNA targets from TarBase [23,24] for each algorithm was compared with the distribution of the full set of miRNA targets predicted by the same algorithm. The overlap of miRNA targets between TarBase [23,24] and each computational algorithm were included in each comparison. For example, the distribution of miRanda-predicted miRNA targets [25,26] was compared with the distribution of targets among them supported by TarBase [23,24].

### ExprTarget: An integrative approach to miRNA target prediction

We constructed a computational prediction model that integrates select miRNA computational algorithms using the experimentally validated targets as training data. Following a multivariate logistic regression model, let the $x_{j,i}$ be the prediction score of algorithm $j$ on the $i$-th miRNA target site prediction:

$$\log it(p_i) = \ln\left(\frac{p_i}{1-p_i}\right) = \beta_0 + \beta_1 x_{1,i} + \ldots + \beta_k x_{k,i}$$

where $p_i$ defines the score for ExprTarget that may be derived from a sigmoidal transformation of a weighted sum of the scores from select computational algorithms or target site features:

$$p_i = \text{sigmoid}\left(\sum \beta_k x_{k,i}\right)$$

$$\text{sigmoid}(s) = \frac{1}{1+\exp(-s)}$$

Pictar [30], TargetScan [27–29], miRanda [25,26], and our HapMap-based expression microarray dataset were selected as providing the $x_{j,i}$, although the approach we describe here may extend to a larger list provided certain assumptions are met. Estimated from the data, each beta $\beta_k$ describes the size of the contribution from a target site feature (encapsulated by the prediction algorithm $k$). The model, as stated, characterizes the relationship between the various predictors and a binary variable, expressed as a probability, showing whether the prediction is experimentally supported or not. A high absolute magnitude for the parameter $\beta_k$ implies that the incorporation of the respective target prediction feature increases the probability of experimental support. The PicTar [30] score, which has a maximum = 1000, has been normalized to lie between 0 and 1 by dividing by this maximum value. The score for the expression-corroborated miRNA targets is set to the p value for the general linear model between miRNA and mRNA normalized ($\log_2$-transformed) expression intensities, provided the estimated coefficient is negative; otherwise, the expression-based score is set to 1. For the miRanda [25,26] contribution, we used the algorithm's *Score*, which is based on complementary base pairing as well as the presence of mismatches, gap-opening, and gap-extension. For TargetScan [27–29], we utilized the probability of preferentially conserved targeting ($P_{CT}$), a Bayesian estimate of the probability of site conservation due to selective maintenance of miRNA targeting. Alternatively, one could utilize the TargetScan context score, which is meant to provide complementary information to $P_{CT}$ and is derived from information orthogonal to site conservation [27–29]; however, our analysis shows that the use of context scores does not lead to higher predictive performance based on AUC (see next section). The change in ExprTarget score $p$ with respect to the change in score $x_i$ for a given prediction method is given by:

$$\frac{\partial p}{\partial x_i} = p(1-p)\beta_i$$

## Performance evaluation using ROC curve analysis and cross-validation

The various prediction algorithms were assessed using the receiver operating characteristic (ROC) curves, which plot the true positive rate (sensitivity) and the false positive rate (1-specificity) for a binary classifier at various score thresholds [39]. The performance of a prediction algorithm can be evaluated using the area under the ROC curve (AUC). The area estimate $\Omega$ has the following standard error (SE):

$$SE(\Omega) = \left(\frac{\theta(1-\theta) + (n_A-1)(Q_1-\theta^2) + (n_N-1)(Q_2-\theta^2)}{n_A n_N}\right)^{1/2}$$

where $n_A$ and $n_N$ are the number of "failures" and "successes", $\theta$ is the "true" area, and $Q_1$ and $Q_2$ are distribution specific quantities [40]:

$$Q_1 = \frac{\theta}{2-\theta}$$

$$Q_2 = \frac{2\theta^2}{1+\theta}$$

Two prediction algorithms with AUC values $\Omega_1$ and $\Omega_2$ may be compared. We can determine the standard error for the difference in areas as follows [41]:

$$SE(\Omega_1 - \Omega_2) = (SE^2(\Omega_1) + SE^2(\Omega_2) - 2rSE(\Omega_1)SE(\Omega_2))^{\frac{1}{2}}$$

where $r$ is the estimated correlation between $\Omega_1$ and $\Omega_2$. The test statistic for the comparison of areas is defined as follows:

$$Z = \frac{\Omega_1 - \Omega_2}{SE(\Omega_1 - \Omega_2)} \sim N(0,1)$$

Since the experimentally supported targets in TarBase may represent biases (e.g., relative to sequence conservation), we chose to evaluate the performance of each algorithm (miRanda [25,26], Pictar [30] and TargetScan [27–29]) using TarBase [23,24] first as gold standard followed by using TarBase [23,24] and expression-corroborated predictions (TarBase + LCL) as gold standard. We also evaluated the predictive performance of ExprTarget using as gold standard the subset of TarBase [23,24] that includes only the target sites validated by individual experiments rather than by high-throughput assays.

To further assess the predictive power of our integrative approach, we conducted $K$-fold cross-validation on the training algorithm in ExprTarget. This method partitions the entire dataset into $K$ subsets or folds $F_i$, $i = 1,2,…K$. One of the $K$ subsets is used as a validation set, $F_{test}$ where $test \in \{1,2,…,K\}$, and the remaining $K-1$ subsets are combined into a training set $F_{train}$ to fit our weighted logistic regression model. The training algorithm is repeated $K$ times so that each subset is used as a validation set once, and an average error is calculated. All observed data are thus used in the training and the validation. We utilized the *cv.glm* function for cross-validation in generalized linear models in the *boot* package available in R [38]. For $K = 10$, the mean error was 0.00027725 while for $K = 3$, the mean error was 0.00027719.

## Supporting Information

**Figure S1** Individual performance of foundational prediction algorithms using (TarBase + LCL) as gold standard. The three prediction algorithms were evaluated using ROC curve analysis, using (TarBase + LCL) as benchmark.
Found at: doi:10.1371/journal.pone.0013534.s001 (2.30 MB TIF)

**Figure S2** When the subset of TarBase that excludes the high-throughput assays is used as gold standard, the improvement in predictive performance for ExprTarget relative to the individual methods continues to hold robustly.
Found at: doi:10.1371/journal.pone.0013534.s002 (1.02 MB TIF)

**Figure S3** The use of a score from the expression data that is gene-based rather than target site-based (e.g., the score is defined as the minimum of all p values for miRNA correlations with the gene) shows that the incorporation of the individual algorithms improves predictive performance.
Found at: doi:10.1371/journal.pone.0013534.s003 (1.35 MB TIF)

## Acknowledgments

The authors would like to thank Wasim K. Bleibel for his excellent technical support in preparing samples for miRNA analysis and Dr. Stephanie Huang for some helpful discussion.

## Author Contributions

Conceived and designed the experiments: ERG YAL NJC MED WZ. Performed the experiments: ERG SD. Analyzed the data: ERG HKI. Contributed reagents/materials/analysis tools: ERG HKI SD WZ. Wrote the paper: ERG HKI YAL MED WZ.

## References

1. Sotiriou C, Pusztai L (2009) Gene-expression signatures in breast cancer. N Engl J Med 360: 790–800.
2. Nannini M, Pantaleo MA, Maleddu A, Astolfi A, Formica S, et al. (2009) Gene expression profiling in colorectal cancer using microarray technologies: results and perspectives. Cancer Treat Rev 35: 201–209.
3. Bacher U, Kohlmann A, Haferlach T (2009) Perspectives of gene expression profiling for diagnosis and therapy in haematological malignancies. Brief Funct Genomic Proteomic 8: 184–193.
4. Seo D, Ginsburg GS, Goldschmidt-Clermont PJ (2006) Gene expression analysis of cardiovascular diseases: novel insights into biology and clinical applications. J Am Coll Cardiol 48: 227–235.
5. Xu B, Karayiorgou M, Gogos JA MicroRNAs in psychiatric and neuro-developmental disorders. Brain Res 1338: 78–88.
6. The International HapMap Consortium (2003) The International HapMap Project. Nature 426: 789–796.
7. The International HapMap Consortium (2005) A haplotype map of the human genome. Nature 437: 1299–1320.
8. Morley M, Molony CM, Weber TM, Devlin JL, Ewens KG, et al. (2004) Genetic analysis of genome-wide variation in human gene expression. Nature 430: 743–747.
9. Stranger BE, Forrest MS, Dunning M, Ingle CE, Beazley C, et al. (2007) Relative impact of nucleotide and copy number variation on gene expression phenotypes. Science 315: 848–853.
10. Zhang W, Duan S, Kistner EO, Bleibel WK, Huang RS, et al. (2008) Evaluation of genetic variation contributing to differences in gene expression between populations. Am J Hum Genet 82: 631–640.
11. Spielman RS, Bastone LA, Burdick JT, Morley M, Ewens WJ, et al. (2007) Common genetic variants account for differences in gene expression among ethnic groups. Nat Genet 39: 226–231.
12. Stranger BE, Nica AC, Forrest MS, Dimas A, Bird CP, et al. (2007) Population genomics of human gene expression. Nat Genet 39: 1217–1224.
13. Storey JD, Madeoy J, Strout JL, Wurfel M, Ronald J, et al. (2007) Gene-expression variation within and among human populations. Am J Hum Genet 80: 502–509.
14. Stranger BE, Forrest MS, Clark AG, Minichiello MJ, Deutsch S, et al. (2005) Genome-wide associations of gene expression variation in humans. PLoS Genet 1: e78.
15. Zhang W, Dolan ME (2009) Use of cell lines in the investigation of pharmacogenetic loci. Curr Pharm Des 15: 3782–3795.
16. Welsh M, Mangravite L, Medina MW, Tantisira K, Zhang W, et al. (2009) Pharmacogenomic discovery using cell-based models. Pharmacol Rev 61: 413–429.
17. Gamazon ER, Huang RS, Cox NJ, Dolan ME (2010) Chemotherapeutic drug susceptibility associated SNPs are enriched in expression quantitative trait loci. Proc Natl Acad Sci U S A 107: 9287–9292.
18. He L, Hannon GJ (2004) MicroRNAs: small RNAs with a big role in gene regulation. Nat Rev Genet 5: 522–531.
19. Lim LP (2005) Microarray analysis shows that some microRNAs downregulate large numbers of target mRNAs. Nature 433: 769–773.
20. Bushati N, Cohen SM (2008) MicroRNAs in neurodegeneration. Curr Opin Neurobiol 18: 292–296.
21. Barringhaus KG, Zamore PD (2009) MicroRNAs: regulating a change of heart. Circulation 119: 2217–2224.
22. Medina PP, Slack FJ (2008) microRNAs and cancer: an overview. Cell Cycle 7: 2485–2492.
23. Sethupathy P, Corda B, Hatzigeorgiou AG (2006) TarBase: A comprehensive database of experimentally supported animal microRNA targets. Rna 12: 192–197.
24. Papadopoulos GL, Reczko M, Simossis VA, Sethupathy P, Hatzigeorgiou AG (2009) The database of experimentally supported targets: a functional update of TarBase. Nucleic Acids Res 37: D155–158.
25. Enright AJ, John B, Gaul U, Tuschl T, Sander C, et al. (2003) MicroRNA targets in Drosophila. Genome Biol 5: R1.
26. Griffiths-Jones S, Saini HK, van Dongen S, Enright AJ (2008) miRBase: tools for microRNA genomics. Nucleic Acids Res 36: D154–158.
27. Lewis BP, Burge CB, Bartel DP (2005) Conserved seed pairing, often flanked by adenosines, indicates that thousands of human genes are microRNA targets. Cell 120: 15–20.
28. Grimson A, Farh KK, Johnston WK, Garrett-Engele P, Lim LP, et al. (2007) MicroRNA targeting specificity in mammals: determinants beyond seed pairing. Mol Cell 27: 91–105.
29. Friedman RC, Farh KK, Burge CB, Bartel DP (2009) Most mammalian mRNAs are conserved targets of microRNAs. Genome Res 19: 92–105.
30. Krek A, Grun D, Poy MN, Wolf R, Rosenberg L, et al. (2005) Combinatorial microRNA target predictions. Nat Genet 37: 495–500.
31. Bentwich I (2005) Prediction and validation of microRNAs and their targets. FEBS Lett 579: 5904–5910.
32. Sethupathy P, Megraw M, Hatzigeorgiou AG (2006) A guide through present computational approaches for the identification of mammalian microRNA targets. Nat Methods 3: 881–886.
33. Zhang W, Duan S, Bleibel WK, Wisel SA, Huang RS, et al. (2009) Identification of common genetic variants that account for transcript isoform variation between human populations. Hum Genet 125: 81–93.
34. Nicolae DL, Gamazon E, Zhang W, Duan S, Dolan ME, et al. (2010) Trait-associated SNPs are more likely to be eQTLs: annotation to enhance discovery from GWAS. PLoS Genet 6: e1000888.
35. Gamazon ER, Zhang W, Konkashbaev A, Duan S, Kistner EO, et al. (2009) SCAN: SNP and copy number annotation. Bioinformatics 26: 259–262.
36. Klein TE, Chang JT, Cho MK, Easton KL, Fergerson R, et al. (2001) Integrating genotype and phenotype information: an overview of the PharmGKB project. Pharmacogenetics Research Network and Knowledge Base. Pharmacogenomics J 1: 167–170.
37. Ritchie ME, Silver J, Oshlack A, Holmes M, Diyagama D, et al. (2007) A comparison of background correction methods for two-colour microarrays. Bioinformatics 23: 2700–2707.
38. R Development Core Team (2005) R: A language and environment for statistical computing. Vienna, Austria: R Foundation for Statistical Computing.
39. Sing T, Sander O, Beerenwinkel N, Lengauer T (2005) ROCR: visualizing classifier performance in R. Bioinformatics 21: 3940–3941.
40. Hanley JA, McNeil BJ (1982) The meaning and use of the area under a receiver operating characteristic (ROC) curve. Radiology 143: 29–36.
41. Hanley JA, McNeil BJ (1983) A method of comparing the areas under receiver operating characteristic curves derived from the same cases. Radiology 148: 839–843.